# A Review on Integration of Artificial Intelligence and Medical Imaging in IVF Ovarian Stimulation

Jana Zakall [1,+], Birgit Pohn[1,+], Antonia Graf[1], Daniel Kovatchki [2], Arezoo Borji [3,4,5], S M Ragib Shahriar Islam [4,3,5], Hossam Haick [6,7], Heinz Strohmer [2], Sepideh Hatamikia [3,4*]

1 Department of Medicine, Danube Private University (DPU), Krems, Austria
2 Kinderwunschzentrum an der Wien, Austria
3 Research Center for Clinical AI-Research in Omics and Medical Data Science (CAROM), Department of Medicine, Danube Private University (DPU), Krems, Austria
4 Austrian Center for Medical Innovation and Technology (ACMIT), Wiener Neustadt, Austria
5 Department of Medical Physics and Biomedical Engineering, Medical University of Vienna (MUV), Vienna, Austria
6 Laboratory for Life Sciences and Technology (LiST), Department of Medicine, Danube Private University, Krems, Austria
7 Department of Chemical Engineering, Technion – Israel Institute of Technology, Haifa, Israel
+ These authors contributed equally to this work

## Abstract

Artificial intelligence (AI) has emerged as a powerful tool to enhance decision-making and optimize treatment protocols in in vitro fertilization (IVF). In particular, AI shows significant promise in supporting decision-making during the ovarian stimulation phase of the IVF process. This review evaluates studies focused on the applications of AI combined with medical imaging in ovarian stimulation, examining methodologies, outcomes, and current limitations.

Our analysis of 13 studies on this topic reveals that, while AI algorithms demonstrated notable potential in predicting optimal hormonal dosages, trigger timing, and oocyte retrieval outcomes, the medical imaging data utilized predominantly came from two-dimensional (2D) ultrasound which mainly involved basic quantifications, such as follicle size and number, with limited use of direct feature extraction or advanced image analysis techniques. This highlights an underexplored opportunity where advanced image analysis approaches, such as deep learning, and more diverse imaging modalities, like three-dimensional (3D) ultrasound, could unlock deeper insights.

Additionally, the lack of explainable AI (XAI) in most studies raises concerns about the transparency and traceability of AI-driven decisions—key factors for clinical adoption and trust. Furthermore, many studies relied on single-center designs and small datasets, which limit the generalizability of their findings.

This review highlights the need for integrating advanced imaging analysis techniques with explainable AI methodologies, as well as the importance of leveraging multicenter collaborations and larger datasets. Addressing these gaps has the potential to enhance ovarian stimulation management, paving the way for efficient, personalized, and data-driven treatment pathways that improve IVF outcomes.

# 1 Introduction

In vitro fertilization (IVF) is a widely used assisted reproductive technology designed to help individuals and couples overcome infertility [1]. It involves several stages, from ovarian stimulation to oocyte retrieval, fertilization, and embryo transfer [1–3]. Among these, ovarian stimulation is critical as it ensures the development of multiple oocytes, enhancing the chances of successful fertilization and pregnancy [2,4]. However, the process requires precise management, including the selection of stimulation protocols, hormone dosages, and optimal timing for oocyte retrieval, making it a complex and often subjective task. Figure 1 shows the ovarian stimulation process in IVF, comprising four stages: hormonal medication to stimulate follicles, monitoring follicle growth via ultrasound and hormone levels, follicle puncture for oocyte retrieval, and egg retrieval using ultrasound-guided aspiration to collect mature eggs for fertilization.

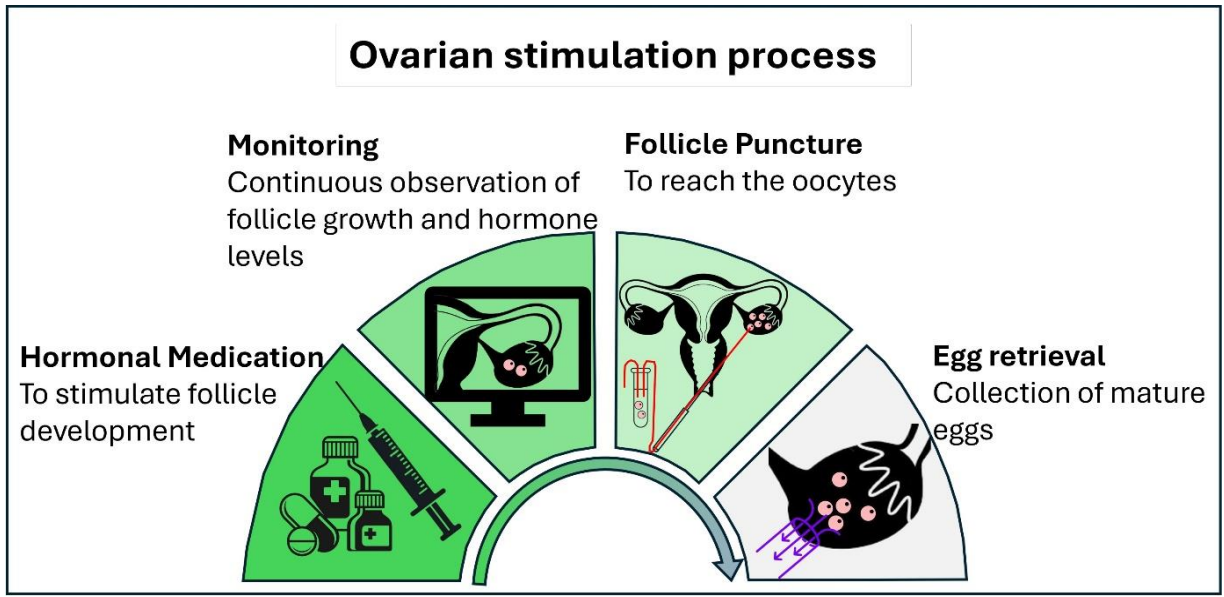

Figure 1 - The Process of Ovarian Stimulation

Currently, ovarian stimulation protocols often follow a "one-size-fits-all" approach. These protocols rely heavily on clinical guidelines and the subjective judgment of clinicians rather than on personalized, data-driven insights [5,6]. While this approach provides general frameworks for treatment, it fails to account for individual patient characteristics. Consequently, the outcomes of ovarian stimulation can vary significantly, with potential under- or over-responses to stimulation, increasing risks of ovarian hyperstimulation syndrome (OHSS) or suboptimal egg retrieval.

The advent of artificial intelligence (AI) offers an opportunity to revolutionize the decision-making process in ovarian stimulation by integrating large datasets and identifying personalized treatment strategies [1,3,7]. Recent studies have explored AI's potential in optimizing hormone dosages, predicting optimal trigger timing for oocyte retrieval, and scheduling ultrasound monitoring [2,4,8–17] which used AI for the goals mentioned here]. However, most AI models have primarily relied on clinical data, such as patient demographics, hormonal profiles, and treatment history, with limited incorporation of medical imaging data [2,4,9,11,13,18–26]. Ultrasound imaging plays a crucial role in IVF procedures, particularly in assessing ovarian and uterine status [10,16,17,27–29]. Before stimulation, pre-stimulation ovarian ultrasounds evaluate baseline ovarian reserves by counting antral follicles (3–8 mm in size). During stimulation, ovarian and uterus ultrasounds monitor follicular growth, measure follicle sizes and numbers, and assess endometrial thickness—critical parameters for adjusting hormone dosages and timing oocyte retrieval. A uterus ultrasound evaluates endometrial thickness, a predictor of implantation success, where a thickness below 6 mm is often associated with reduced pregnancy chances. Despite the routine acquisition of ultrasound images in clinical practice, AI applications have primarily focused on numerical data derived from ovaries images, such as follicle count and size, rather than analyzing the images themselves [2,4,9,11,13,18–26]. Additionally, radiomics features have been utilized in conjunction with AI in only a few studies, and only when analyzing uterine ultrasound data [14,26,10,30]. However, the integration of advanced image analysis techniques, such as deep learning and other sophisticated methods, remains largely underutilized in AI applications for ovarian stimulation.

This review provides a comprehensive overview of studies utilizing AI and medical imaging to support ovarian stimulation decision-making. It highlights the current research landscape, emphasizing the integration of imaging data with clinical parameters, and identifies gaps in leveraging imaging technologies. By addressing these gaps, this paper aims to provide recommendations for future advancements in AI-driven personalized treatment strategies for ovarian stimulation, ultimately improving IVF outcomes.

## 2  Materials and Methods

### 2.1  Search strategy and eligibility criteria

This review paper aims at analyzing the publications which used AI and medical imaging data to support treatment decision making for ovarian stimulation process in IVF. To identify relevant literature comprehensively, a PRISM analysis was conducted across multiple academic databases, including FertStert, PubMed Central, Elsevier/Embase, IEEE Xplore and MDPI with targeted search terms such as "IVF", "in vitro fertilization", "reproductive medicine" in combination with "artificial intelligence", "machine learning", "deep learning", "fertility", "imaging," "ultrasound," "MRI," and "radiomics". Synonyms and alternative terms were included to broaden the scope and ensure a thorough capture of studies.

## 2.2 Procedure of the PRISM analysis

Following the initial search, duplicates were removed using reference management software Zotero to retain unique study records. The remaining articles were screened in two stages: by title and abstract, and then by full-text review, with inclusion criteria focusing on studies using AI and medical imaging in IVF treatments. Exclusions were documented with specific reasons, providing transparency in the selection process. An initial pool of 3,344 records was identified. After removing duplicates, the remaining papers for relevance to IVF treatment variables involving stimulation, follicle size, dosage, and trigger timing, excluding studies on non-human data and male infertility were screened. This filtering reduced the dataset to 54 papers. Next, we excluded studies focusing on embryos, endometrium/uterus only, or sperm, along with reviews and collection papers. Finally, we restricted the selection to studies that utilized AI and medical imaging data focusing on ovarian stimulation, resulting in 13 studies that were ultimately included for our analysis.

# 3 Results

Publications selected according to the search criteria (Section 2.1) were categorized according to several aspects, including the target of the study, the imaging modality used, the integration of imaging and other data types, the AI methods used, the number of patients used, the use of multiple centers or only one center, and the explainable AI used. Details of each categorization can be found below. Tables 1 and 2 summarize all details related to all different categorizations.

## 3.1 Target of the selected studies and their performance

### 3.1.1 Hormonal dose optimization and optimal trigger date

In the study from Gerard Letterie et al. [13], the authors proposed an AI algorithm and assessed its accuracy for the day-to-day decision making during ovarian stimulation compared to evidence-based decisions by the clinical team. Their AI method showed varying accuracy levels for different goals i.e., 0.82 for medication dose adjustments, 0.92 for deciding whether to continue or stop treatment, 0.96 for scheduling or canceling oocyte retrieval and 0.87 for determining the number of days until the next follow-up. Their proposed algorithm was quite accurate in making decisions about whether to continue or stop treatment and whether to trigger egg release or cancel the cycle. However, it wasn't as accurate when deciding on optimal hormonal medication doses as the algorithm was not often able to predict the change in dose when an increase was needed.

In the study by Eduardo Hariton et al. [11], the authors designed an AI algorithm to decide the best day to trigger egg retrieval in IVF cycles. When focused on maximizing the number of fertilized eggs, the algorithm recommended triggering 42.3% of the time and waiting 57.7% of the time. Following the algorithm's recommendations resulted in an average gain of 3.015 more fertilized eggs. When the algorithm was adjusted to optimize the number of usable blastocysts (early-stage embryos), it suggested triggering 45.2% of the time and waiting 54.8% of the time. Although their approach led to an average gain of 1.430 more fertilized eggs and 0.577 more total usable blastocysts the algorithm could effectively identify the best trigger day, leading to better outcomes in terms of fertilized eggs and usable blastocysts.

The study of Abdel Hameed et al. [10] aimed to develop an AI model to predict oocyte retrieval during controlled ovarian stimulation (COS) and create user-friendly nomograms to guide gonadotropin protocol and dose decisions.

Michael Fanton et al. [4] developed interpretable AI models to predict mature (MII) oocyte outcomes if ovulation was induced on the current day versus the following day. Linear regression models aimed to predict MII oocyte, while an estradiol (E2) forecasting model was designed to predict next-day E2 levels, both models using prior follicle counts and E2 data. These models allowed a comparative analysis of MII outcomes between potential trigger days. The average trigger day for ovulation was reported around day 11.8, with patients having an average of 4.5 monitoring visits per cycle. Their model predicting MII oocytes on the trigger day achieved a mean absolute error (MAE) of 2.87 oocytes and an $R^2$ of 0.64, while the prediction model for next-day MII outcomes had an MAE of 3.02 oocytes and an $R^2$ of 0.62. For next-day estradiol (E2) levels, the prediction model performed better, with an MAE of 274 pg/mL and an $R^2$ of 0.88. Adding a follicle imputation algorithm improved the model's accuracy slightly, enhancing the MAE by 0.09 oocytes and $R^2$ by 0.02. The study shows an effective way of creating an interpretable machine learning model to optimize the day of trigger.

The study by Gerard Letterie et al. [2] aimed to simplify the IVF process using AI by predicting the best monitoring and trigger dates while maximizing the number of retrieved oocytes. Their algorithm was able to predict the optimal monitoring day during ovarian stimulation with a mean absolute error (MAE) of 1.355. For prediction of the best trigger day, the predicted number of oocytes varied minimally (0–3 oocytes) across these three days, showing minimal impact from slight shifts in the trigger date. Their algorithm predicted total and mature oocyte numbers with an MAE of 3.517. Sensitivity for predicting mature oocytes was higher for patients with more than 10 oocytes (0.81) than for those with 0–10 oocytes (0.78). Anti-Müllerian hormone (AMH) was the most influential predictor for accuracy.

Xiaowen Liang et al. [9] addressed two different aspects in his study: firstly, how human chorionic gonadotropin (HCG) administration can be optimized through using 3D ultrasound to assess follicle volume and a deep learning-based biomarker method, and secondly, how their deep learning follicle volume biomarker can predict the number of mature oocytes. The second aspect is explained in the next section. The optimal leading follicle volume as a threshold for HCG timing was 3.0 cm³, showing a statistically significant association with higher maturity. The HCG trigger (1000–2000 IU) was administered when at least three 17 mm follicles or two 18 mm follicles were visible in the ultrasound. The findings confirmed that using the 3.0 cm³ follicle volume as a biomarker could reliably guide the optimal timing of the HCG administration, being an improved method for achieving optimal oocyte maturity before retrieval.

### 3.1.2 Oocyte quality

In the second aspect explored by Xiaowen Liang et al.'s study [9], a 3D deep learning biomarker, which automatically calculates follicle volume, and the number of retrieved mature oocytes were compared with 2D follicular diameter measurements. Results showed that an optimal leading follicle volume threshold for prediction of mature oocytes retrieved was 0.5 cm3 or larger on the HCG administration day. Furthermore, the study found that the 3D-US follicle marker improved retrieval outcome in comparison with traditional 2D measurements.

The study of Pedro Royo et al. [31] OSIS Ovary by comparing its performance with the traditional 2D manual measurement method. OSIS Ovary is a 3D ultrasound system for automated follicle tracking during ovarian stimulation and was designed to automate follicle segmentation, visualization and measurement. The system demonstrated high reliability with strong correlation coefficients (≥0.9 for follicle diameter and ≥0.8 for follicle count) compared to manual measurements across different follicle sizes (≥10mm, ≥13mm, and ≥16mm). Slight differences in mean diameter (>1mm) and steady limits of agreement ranges (<6mm) were observed, making OSIS comparable to manual methods. Furthermore, OSIS Ovary performed similarly to SonoAVC™, a known automated system, with slightly shorter limits of agreement and minimal deviations. OSIS Ovary is a promising and reliable tool for tracking follicle growth during ovarian stimulation.

Fangfang Xu et al. [14] developed a robust, explainable machine learning model combining radiomics and clinical information to forecast frozen embryo transfer (FET) outcomes. To do this, radiomics data, derived from ultrasound images of the endometrium and the junctional zone, was used to calculate a rad-score. Clinical factors like patient age, endometrial thickness (EMT), and embryo quality were analyzed separately. An integrated model, fusion model A, was created by combining radiomics and clinical data. Fusion model A demonstrated the best accuracy for predicting pregnancy outcomes, with an AUC of 0.861 in training and 0.793 in testing. It outperformed models based solely on radiomics or clinical data. The most influential factors in predicting outcomes were the rad-score, embryo grade, patient age, and EMT. Higher rad-scores and better embryo grades were associated with higher success rates.

The study of Michael Fanton et al. [32] aimed to evaluate the integration of two AI tools: MyCycleClarity for automated follicle measurement and Alife Stim Assist™ for predicting egg retrieval outcomes, comparing their accuracy to human measurements. Follicles sized 14–17mm showed the strongest association with egg retrieval outcomes. MyCycleClarity identified significantly more small follicles (<10mm) than manual counts, but similar numbers of large follicles (>10mm). Stim Assist™ predictions using AI measurements had a smaller mean absolute error (3.30 eggs) than those using human measurements (3.84 eggs). Integrating MyCycleClarity and Stim Assist™ slightly improves prediction accuracy for egg retrieval compared to human follicle measurements, likely due to more comprehensive follicle counting by AI, especially for smaller follicles.

Hanassab S. et al. [15] used ensemble-based explainable artificial intelligence (XAI) to find out which follicle sizes on trigger day (TD) are most critical for producing mature oocytes, embryos, and blastocysts, and improving live birth rates. Follicles with 13–18mm on TD were most likely to yield mature oocytes, particularly 15–18mm, those sized 14–20mm contributed to high-quality blastocysts. Follicles with 12–19mm were optimal in gonadotropin-releasing hormone (GnRH) antagonist protocols, follicles with 14–20mm in GnRH agonist cycles. On TD, having three leading follicles ≥17mm improved mature oocyte yield by 10%. Optimizing the proportion of follicles sized 13–18mm on TD significantly enhances live birth rates

### 3.1.3 Endometrium receptivity

Wendi Huang et al. [28] explored the use of radiomics features to assess endometrial receptivity (ER) in patients with recurrent pregnancy loss (RPL). A radiomics score (rad-score) was developed using features extracted from ultrasound images. Five significant radiomics signatures were identified, forming the rad-score which was strongly associated with RPL. The rad-score outperformed traditional predictors like age, spiral artery pulsatility index (SA-PI), and vascularization index (VI) in distinguishing RPL cases. Increased rad-scores (age, SA-PI, and SA-RI) were observed in RPL patients, while VI was lower compared to controls. The study found that rad-scores provide a promising tool for predicting ongoing pregnancy potential in RPL patients, complementing existing ER evaluation methods.

Another prospective cohort study performed by Xiaowen Liang et al. [16] evaluated if a multi-modal fusion model combining ultrasound-based deep learning radiomics features and clinical parameters can predict clinical pregnancy outcomes after frozen embryo transfer (FET). The model's performance was evaluated using metrics such as area under the curve (AUC), accuracy, sensitivity, and specificity, with the proposed model achieving an AUC of 0.825. Key findings indicated that the multi-modal fusion model outperformed models using either clinical or imaging data alone, demonstrating the potential of deep learning in enhancing predictive accuracy for FET outcomes.

The study of J. Fjeldstad et al. [33] explored if an AI-based model is able to predict endometrial receptivity for successful embryo implantation using ultrasound images taken during ovarian stimulation. Implantation outcomes were linked to blastocyst quality and genetic testing. Three model training scenarios based on blastocyst quality were tested, and performance was measured using AUC, sensitivity, and specificity. The best-performing model, trained on all blastocysts (scenario 3), achieved AUC 0.631, sensitivity 0.628, and specificity 0.556. Clinical features like endometrial thickness, progesterone levels, patient age, and embryo transfer history were significant predictors. Endometrial thickness alone at its best threshold (8.8 mm) achieved lower predictive power (AUC 0.576). Therefore, an AI model combining ultrasound imaging and clinical data predicts implantation success more accurately than endometrial thickness (EMT) alone.

Table 1 summarizes key study details, including publication year, study targets, imaging techniques used (e.g., 2D/3D ultrasound), specific imaging data analyzed, and additional clinical or data modalities integrated.

Table 1 - Overview on target of studies and used (imaging) data

| Ref# | Year of publication | Target of the study | Imaging data used | Information of imaging data used | Other data used |
|---|---|---|---|---|---|
| [4] | 2022 | To develop an interpretable model for predicting the trigger day during ovarian stimulation. | 2D Ultrasound of ovaries | Follicle diameters | Age, BMI, previous IVF cycles, AMH, AFC, estradiol |
| [9] | 2022 | To determine whether follicle volume biomarkers from 3D-US can predict oocyte maturity and optimize HCG trigger timing. | 3D Ultrasound of ovaries | Follicle volumes | Infertility workup, hormonal analysis |
| [10] | 2022 | To identify key follicle sizes on the trigger day for optimizing mature oocytes, embryos, and live birth rates. | 2D Ultrasound of ovaries | Follicle measures | Estradiol, AMH levels, basal FSH, age, cycle data |
| [11] | 2021 | To optimize the trigger day for maximizing fertilized oocytes and blastocyst yield using a machine learning model. | 2D Ultrasound ovaries and endometrium | Endometrial thickness, Follicle sizes (16–20 mm, 11–15 mm) | Cycle data, patient age, BMI, estradiol |
| [13] | 2020 | To assess the accuracy of a computer algorithm in managing day-to-day decisions during ovarian stimulation. | 2D Ultrasound ovaries | Follicle diameters | Estradiol concentrations, FSH dosage, cycle day |
| [14] | 2024 | To predict reproductive outcomes after frozen embryo transfer using a machine learning model combining radiomics and clinical data. | 2D Ultrasound of endometrium and junctional zone | Radiomics features | Clinical data, hormone levels, patient history |

| Ref | Year | Objective | Imaging | Features | Clinical data |
|---|---|---|---|---|---|
| [15] | 2024 | To determine optimal follicle sizes on the trigger day for maximizing mature oocyte yield. | 2D Ultrasound ovaries | Follicle sizes | Fertilization rates, high-quality blastocyst development, live birth rates |
| [16] | 2023 | To predict clinical pregnancy after frozen embryo transfer using a multi-modal AI model combining ultrasound and clinical data. | 2D Ultrasound Endometrium | Radiomics features | serum hormone concentration: oestradiol, progesterone, LH, testosterone, prolactin, anti-Müllerian hormone |
| [21] | 2022 | To automate workflow and enhance decision-making during ovarian stimulation through AI. | 2D Ultrasound of ovaries | Follicle diameters | FSH dosage, estradiol, AMH, BMI, previous IVF cycles |
| [28] | 2023 | To identify radiomics features predicting endometrial receptivity in recurrent pregnancy loss patients. | 2D Ultrasound Endometrium | Radiomics features | Age, BMI, prior miscarriages, hormone profile (FSH, LH, E2), AMH |
| [31] | 2024 | To compare 3D ultrasound-based automated systems with 2D manual methods for tracking follicle growth. | 3D ultrasound ovaries | Follicle sizes, counts | None |
| [32] | 2023 | To evaluate the integration of AI tools for predicting ovarian stimulation outcomes. | 3D ultrasound of ovaries | Follicle sizes, measured manually and via AI | Estradiol, follicle cycle data |
| [33] | 2024 | To develop an AI model predicting endometrial receptivity for embryo implantation using ultrasound and clinical data. | 2D Ultrasound of endometrium | Endometrial thickness, radiomic features | progesterone levels, patient age, embryo transfer history, and blastocyst quality |

Table 2 outlines AI methods, accuracy results, patient numbers or cycles, study types (single or multicenter), and whether explainable AI (XAI) techniques were employed in the analysed studies.

Table 2 - Overview AI methods and results

| Ref# | AI method used | Results | Patient number or cycles | Single or multicenter | XAI |
|---|---|---|---|---|---|
| **[4]** | Linear regression, follicle imputation algorithm | MII prediction: MAE 2.87, R² 0.64; E2 levels: MAE 274 pg/mL, R² 0.88 | 30,278 cycles | Multicenter | Y |
| **[9]** | Deep learning segmentation (C-Rend) | HCG timing threshold: 3.0 cm³; retrieval threshold: 0.5 cm³; 3D-US outperformed 2D | 515 IVF cycles | Single | N |
| **[10]** | Nomogram, Poisson model with log-link function | the model showed superior precision and performance (λ=8.27; relative standard error (λ)=2.02%) | 636 patients | Single | N |
| **[11]** | T-Learner, LightGBM | Trigger timing improved outcomes: +3.015 fertilized eggs, +1.515 usable blastocysts | 7,866 cycles | Single | N |
| **[13]** | Regression trees, random forests, Support Vector Machine (SVM), neural networks | Accuracy: 0.82 for medication, 0.92 for treatment continuation, 0.96 for scheduling retrieval, 0.87 for follow-ups | 2,603 cycles | Single | N |
| **[14]** | Radiomics, XGBoost | AUC 0.861 (training), 0.793 (testing); rad-score, embryo grade, and EMT key predictors | 787 patients | Single | Y |
| **[15]** | Ensemble-based XAI model | Follicles 13–18mm yielded +42% oocytes; LBR increased to 31.6% for optimal follicle proportions | 19,082 patients | Multicenter | Y |
| **[16]** | Deep learning, XGBoost | AUC 0.825; multi-modal fusion model improved pregnancy prediction accuracy | 326 patients | Single | N |
| **[21]** | Stacking ensemble model (linear regression, KNN, XGBoost) | Monitoring day MAE 1.355; trigger day MAE 3.517; sensitivity 0.81 for >10 oocytes | 1,591 cycles | Single | N |
| **[28]** | Radiomics analysis | Rad-score outperformed traditional predictors (SA-PI, VI); linked to higher pregnancy success | 535 patients (262 RPL, 273 controls) | Single | N |
| **[31]** | OSIS Ovary software | strong correlation coefficients (≥0.9 for follicle diameter and ≥0.8 for follicle count) compared to manual measurements across different follicle sizes | 534 patients | Single | N |

| [32] | Linear regression, MyCycleClarity, Stim Assist™ | AI: MAE 3.30 eggs; Human: MAE 3.84 eggs; more small follicles (<10mm) identified by AI | 553 patients | Single | N |
| [33] | Ensemble model (deep learning + feature-based ML) | AUC 0.631 (scenario 3); sensitivity 0.628; specificity 0.556 | 40,910 patients, 79,602 ultrasound images | Multicenter | N |

## 3.2 Imaging data used for supporting decision making of IVF ovarian stimulation process

This section reports on the medical imaging techniques and data used in the searched literature. Gerard Letterie et al. [13] analyzed follicle growth and maturation using transvaginal ultrasound to support decision making for day-to-day management of ovarian stimulation using AI. In this study, follicle diameters in two dimensions in millimeters (mm) for monitoring visits during the stimulation phase of IVF cycles have been measured.

With the aim to investigate whether a ML algorithm can optimize the day of trigger to improve IVF outcomes, Hariton et al. [11] analyzed imaging data such as endometrial thickness gained from transvaginal ultrasound as well as the number of follicles categorized into groups of 5-millimeter (mm) diameter measured at monitoring visits during ovarian stimulation.

Ebid et al. [10] used ultrasound to assess antral follicle development with the aim of creating a validated model for predicting oocyte retrieval outcomes in controlled ovarian stimulation. On the third day of the menstrual cycle, qualified radiographers measured the antral follicles in both ovaries using transvaginal ultrasonography before start of the ovarian stimulation. A follow-up longitudinal assessment of the antral follicle count was also conducted one week after the start of ovarian stimulation to allow adjustments to the gonadotropin dosage.

Fanton et al. [4] analyzed baseline antral follicle count (AFC) from pre-stimulation ultrasound images and follicle sizes measured in monitoring visits during ovarian stimulation. Follicles were categorized into six groups based on their diameter size: <11 mm, 11–13 mm, 14–15 mm, 16–17 mm, 18–19 mm, and >19 mm. They realized that this information could contribute to the development of a ML model for optimizing the day of trigger during ovarian stimulation.

Gerard Letterie et al. [21] included imaging data obtained from ovarian ultrasound during ovarian stimulation. They analyzed the total count of follicles, follicle diameters in two dimensions as the mean value over all follicles as well as outcomes at the time of retrieval including the number of total and mature oocytes, which were categorized into two groups: Group I (0–10 mature oocytes) and Group II (more than 10 mature oocytes). They reported that the information extracted from the follicle diameter combined with AI can support reducing monitoring during ovarian stimulation to a single day and enabling a more balanced scheduling of oocyte retrievals.

In Liang et al.´s study [9] a follicle volume biomarker was developed using a deep learning-based segmentation algorithm to calculate the ideal follicle volume for forecasting the number of mature oocytes retrieved and for fine-tuning the timing of the HCG. For this reason, ovarian ultrasound was used to measure the follicle diameters in mm. All ultrasound follicular monitoring was performed using a 3-dimensional (D) ultrasound device with transvaginal volume probe on the day of HCG administration. The authors reported that the proposed follicle volume biomarker combined with AI can efficiently predict the oocyte maturity in the IVF procedures.

Huang et al. [28] used transvaginal ultrasound data in order to extract radiomics features from ultrasound images of the endometrium to optimize the assessment of endometrial receptivity. The authors reported on the potential of ultrasound radiomics signatures to differentiate between patients with unexplained recurrent pregnancy loss (RPL) and healthy individuals.

Liang et. al [16] used 2D ultrasound images of the endometrium to create a multi-modal fusion model based on ultrasound-based deep learning radiomics for the evaluation of endometrial receptivity and prediction of pregnancy after FET. The authors highlighted the significantly better performance of the multi-modal fusion model compared to using either images or clinical parameters independently for predicting clinical pregnancy outcomes following FET.

Fanton et al. [32] used 3D ultrasound images of the follicles for the assessment of an AI-based tool that can automatically count and measure follicles and predict the eggs retrieved.

The authors reported that the tool to predict the eggs retrieved provided slightly more accurate predictions when using AI-counted follicles compared to human-counted follicles.

Royo et al. [31] used transvaginal 2D and 3D ultrasound in order to evaluate an AI-based solution OSIS Ovary (Online System for Image Segmentation for the Ovary), an automated 3D ultrasound-based system, in comparison with the standard two-dimensional manual measurement method to determine the reliability of measurements of follicle size and count, for tracking follicle growth during ovarian stimulation. A total of three ultrasound scans on three different days of ovarian stimulation, days 4–5 for the initial follicular assessment, days 6–7 for the GnRH antagonist introduction, and days 8–9 and information about follicle size and count was used for oocyte retrieval scheduling.

To find out which follicle sizes on trigger day are essential for receiving mature oocytes, embryos, and blactocysts, ultimately improving live birth rates Hanassab S. et al. [15] follicle diameters obtained by ultrasound scans.

In another study [33], ultrasound images of the endometrium were used to assess a non-invasive AI-based model that evaluates the receptivity of the endometrium to predict the successful implantation of an embryo. The authors of the study stated that the model predicts successful implantation more effectively than endometrial thickness alone.

Fangfang Xu et al. [14] explored the use of ultrasound-derived radiomics features from the endometrium to assess whether an ML model could predict the outcomes of FET.

### 3.3 Integration of imaging and other data types

Most studies which used medical imaging data and AI to support ovarian stimulation process, incorporated a combination of imaging data with other types of information, primarily clinical parameters [2,4,6,7,9,10,11,12,13,15] to derive predictions (Figure 2).

Gerard Letterie et al. [13] included clinical data such as estradiol concentrations in picograms per mm and dose of recombinant follicle-stimulating hormone during ovarian stimulation for IVF in addition to the ovarian ultrasound imaging data in their proposed AI methods.

Eduardo Hariton et al. [11] analyzed clinical parameters such as age, body mass index, protocol type and oestradiol level in addition to the uterus ultrasound imaging data. They reported that the total number of follicles with diameter of 16–20 mm, the total number of follicles with diameter of 11–15 mm, and estradiol level were most important for the high performance of their AI model.

Although Ebid et al. [10] used transvaginal ultrasound to check the development of antral follicles, only clinical data such as women´s age, body mass index, concentrations of basal follicle-stimulating hormone (FSH) measured on day 2–3 of the menstrual cycle, anti-Müllerian hormone levels (AMH), the starting and total gonadotropin dose, duration of the stimulation, and type of the stimulation protocol were included in the prediction model. Women's age, basal FSH levels, antral follicular count (AFC), stimulation protocol type, and gonadotropin dose have been identified as significant predictors of oocyte retrieval, with intrinsic factors such as age and ovarian reserve tests (ORTs) including AMH and FSH levels, and AFC, having the greatest influence.

In addition to the day-by-day follicle measurements from ultrasound imaging data, clinical parameters such as age, BMI, number of previous IVF cycles, baseline AMH levels, baseline E2 levels and cycle length in days have been included in the study from Fanton et al. [4] with the aim of developing an interpretable machine learning model for optimizing the day of trigger during ovarian stimulation.

Along with the follicle measurements from ultrasound imaging Letterie et al. [2] included data obtained both before IVF and during IVF in their study. Pre-IVF data included clinical parameters such as patient age, BMI, smoking history, previous IVF cycles, serum concentrations of AMH and AFC. The data obtained during IVF included protocol type, dose of recombinant FSH, trigger type and estradiol concentrations.

In addition to the follicle volume biomarker obtained from 3D ultrasound data, Liang et al. [9] analyzed a dataset of clinical data from previous treatments and infertility assessments, containing the cause and duration of infertility, BMI, hormonal analysis, stimulation type and dosage, type of trigger and number of mature oocytes retrieved. Liang et al. [16] combined in their study 2D transvaginal ultrasound images, from which they extracted radiomics features, with clinical parameters such as serum hormone concentration, oestradiol, progesterone, LH, testosterone, prolactin and anti-Müllerian hormone.

With the aim of predicting endometrial receptivity in recurrent pregnancy loss patients Huang et al. [28] evaluated along with the radiomics features from ultrasound images, clinical data such as age, BMI, number of prior miscarriages, day-3 hormone profile (FSH, LH, and E2) during the natural menstrual cycle, and AMH rate.

Fanton et al. [32] primarily relied on 3D ultrasound imaging data to measure follicle size and counts. These simple imaging-derived features were combined then with clinical data, specifically estradiol levels, to predict the number of retrieved eggs.

With the aim of detecting the optimal follicle sizes on TG Hanassab S. et al. [15] included follicle measurements from imaging data, as well as clinical parameters such as fertilization rates, high-quality blastocyst development, live birth rates, age and suppressant protocol.

To generate an AI-based endometrial receptivity model Fjeldstad et al. [33] supplemented the ultrasound images of the endometrium with clinical data, including endometrial thickness, progesterone blood levels, age at transfer, the total number of previous embryo transfers, the interval between the ultrasound and transfer dates, age at oocyte retrieval, and the origin of the oocytes.

Fangfang Xu et al. [14] combined image data from 2D and 3D ultrasound images of the endometrium, from which radiomics were extracted, with clinical data such as age, EMT and embryo quality to develop a ML model to forecast FET outcomes.

Figure 2 illustrates data integration in IVF, using inputs like patient biography, hormonal levels, follicle size, ultrasound data (ovary, uterus), and prior IVF results to predict outcomes such as oocyte retrieval, trigger timing, ovarian hyperstimulation risk, and optimal hormonal dosage, leveraging AI.

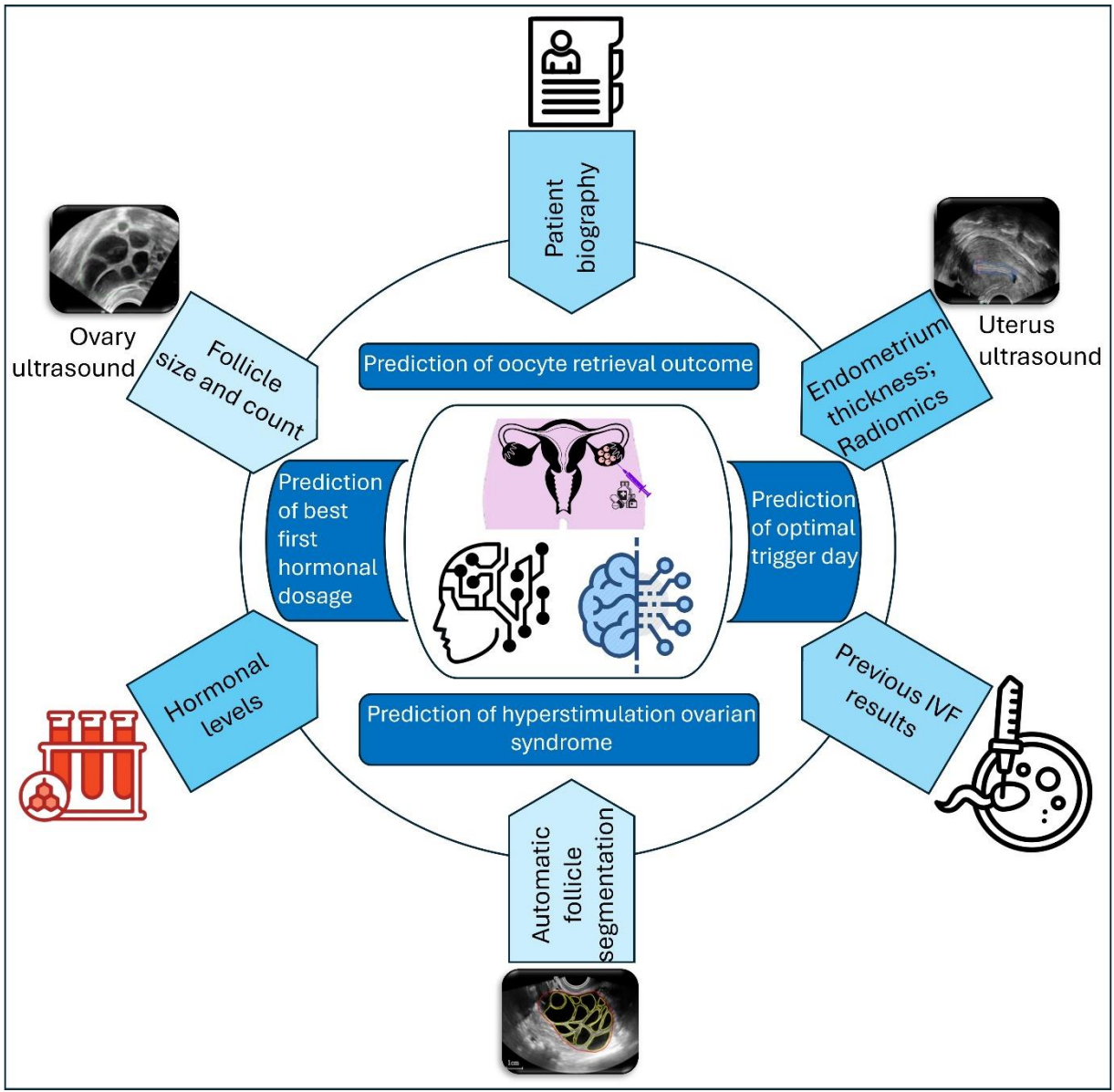

Figure 2 – Integration of different data types i.e., imaging and clinical data to develop AI-based methods to support ovarian stimulation in IVF.

## 3.4 AI algorithm and patient number used

Gerard Letterie et al. [13] employed a hybrid algorithm integrating traditional machine learning methods, including classification and regression trees, random forests, support vector machines (SVM), logistic regression, and neural networks. The algorithm was trained on data from 2,603 IVF cycles (1,853 autologous and 750 donor cycles) comprising 7,376 visits and 59,706 data points. An additional 556 unique cycles were used to validate its performance.

The study of Hariton et al. [11] utilized a ML algorithm based on a T-learner meta-algorithm with bagged Light Gradient Boosting Machines (LightGBM) as base learners. The algorithm analyzed 7,866 IVF cycles to optimize the day of trigger shot administration for improving outcomes like fertilized oocytes (2PNs) and usable blastocysts.

Hameed et al. [10] employed a modified Poisson model with a log-link function to predict oocyte retrieval in a controlled ovarian stimulation (COS) group. The model analyzed data from 636 women undergoing their first in vitro fertilization/intracytoplasmic sperm injection (IVF/ICSI) cycles. The focus was on advanced traditional ML models to establish dose-response relationships and optimize gonadotropin dosing.

The study of Fanton et al. [4] utilized a K-nearest neighbors (KNN) model, a non-linear and interpretable machine learning algorithm, to predict individualized gonadotropin starting doses for ovarian stimulation. The algorithm was trained and tested on 18,591 autologous IVF cycles from three U.S.-based clinics. The KNN model identified the 100 most similar patients to a target individual to create personalized dose–response curves predicting metaphase II (MII) oocytes.

The study of Letterie et al. [2] used an ensemble model combining multiple ML methods (linear regression, random forest, K-nearest neighbor, extra trees regression, and XGBoost) to predict key outcomes, including the optimal day for monitoring, trigger day assignment, and the number of retrieved oocytes. The analysis included data from 1,591 autologous IVF cycles across 4,731 visits, with 80% used for training and 20% as a test set.

Xiaowen Liang et al. [9] utilized a deep learning-based segmentation algorithm to establish a novel follicle volume biomarker for predicting oocyte maturity, involving a total of 515 IVF cases. They also used traditional ML algorithms, specifically a multilayer perceptron (MLP), which demonstrated significant accuracy in predicting ovarian hyper-response. The MLP achieved the best performance, with an accuracy of 0.890 and an AUC of 0.880, indicating its effectiveness in individual predictions. Other multivariate classifiers were also assessed, including decision trees, k-nearest neighbors, random forests, and support vector machines, highlighting the study's comprehensive approach to integrating both deep learning and traditional machine learning techniques in analyzing IVF patient data.

The study of Wendi et al. [28] study involved 600 women, including 300 with unexplained RPL and 300 who had full-term pregnancies without prior loss. Significant features were identified using LASSO (Least Absolute Shrinkage and Selection Operator) and logistic regression, and a radiomics score (rad-score) was calculated. The rad-score outperformed conventional ER indicators in identifying RPL patients.

The study of Liang X. et al. [16] used a total of 326 patients who underwent frozen FET. They proposed a multi-modal fusion model that combined quantitative clinical parameters (tabular data) with two-dimensional ultrasound images. For image analysis, the study utilized a deep learning architecture based on the VGG-11 network to extract radiomic features from ultrasound images, also focusing on the segmentation of the endometrium to identify regions of interest. It was reported that the multi-modal fusion model outperformed models using either clinical or imaging data alone, demonstrating the potential of deep learning in enhancing predictive accuracy for FET outcomes.

Fanton M, Wenchel S. et al. [32] evaluated the integration of two AI tools for ovarian stimulation: MyCycleClarity, an AI tool for automated follicle counting and measurement from 3D ultrasound images (trained on 91,782 follicles across 19,776 ovaries), and Alife Stim Assist Trigger Tool, a linear regression model predicting the number of eggs retrieved (trained on 26,179 cycles). Data from 553 patients at a single U.S. clinic were analyzed, including 82 cycles with human follicle measurements, 186 cycles with AI measurements, and 25 cycles with both methods. On the day of trigger, MyCycleClarity identified significantly more small follicles (<10mm) compared to manual counting (9.5 vs. 0.8 on average) and a similar number of large follicles (>10mm). The Stim Assist Tool achieved a mean absolute prediction error of 3.30 eggs using AI follicle measurements, compared to 3.84 eggs with human measurements, showing slightly higher accuracy with AI-derived data. The results suggest that MyCycleClarity's improved ability to count small follicles enhances the performance of the Stim Assist Tool, highlighting the synergy between AI tools for more accurate predictions of egg retrieval outcomes in ovarian stimulation.

The study from Royo et al. [31] included 89 female participants undergoing ovarian stimulation, generating 534 transvaginal 3D ultrasound datasets. The OSIS Ovary system was used which employed a deep convolutional neural network (DCNN) with residual connections to enhance segmentation accuracy based on a modified U-Net architecture. This DCNN was trained on 100 manually segmented ovarian volumes to perform automatic segmentation of the follicles. The key image features extracted included the number of follicles and their relaxed sphere diameters, which were crucial for evaluating follicle growth.

Hanassab et al. [15] utilized an ensemble-based explainable artificial intelligence (XAI) model combining machine learning methods optimized through Bayesian tuning. The model analyzed data from 19,082 patients undergoing their first IVF/ICSI cycle across a span of 18 years. Permutation importance and SHAP values were applied to interpret key follicle size contributions to the prediction.

The study by Fjeldstad et al. [33] employs an ensemble AI model combining a deep learning image-based component and a feature-based machine learning (ML) model to predict endometrial receptivity for successful embryo implantation. The deep learning model analyzed 79,602 ultrasound images of the endometrium, while the ML model incorporated clinical features such as endometrial thickness, progesterone levels, patient age, embryo transfer history, and blastocyst quality. The dataset comprised 40,910 patients from 70 clinic locations across four clinic networks, providing a robust foundation for model development and validation.

### 3.5 Single or multi-center study

Out of the 13 studies, 10 studies used singlecenter data [9–11,13,14,16,21,28,29,31,32] while 3 used multicenter data [15,20,33]. Single-center studies ensure consistent data collection but are normally limited by smaller sample size and less patient data diversity and thus reduced generalizability. Multicenter studies provide diverse data sources, enhancing robustness but introducing variability in data collection. Larger cohorts, such as those in multicenter studies [15,20, [33]], improve reliability and validation of ML algorithms. Table 3 categorizes studies based on patient or cycle counts: those exceeding 10,000 patients/cycles [15, 20, 33], between 1,000–10,000 [13, 11, 21], and fewer than 1,000 [10, 14, 9, 28, 16, 31, 33]. Most studies fall into the smaller cohort categories, reflecting limited large-scale data.

Table 3 - Overview of Patient Numbers per Study

| Category | Over 10,000 Patients/Cycles | 1,000–10,000 Patients/Cycles | Fewer than 1,000 Patients/Cycles |
| --- | --- | --- | --- |
| **Number of Patients/Cycles** | 19082 [15], 30,278 cycles [20], 40910 [33] | 2603 + 556 cycles [13], 7866 cycles [11], 1591 [21] | 636 [10], 787 [14], 515 [9], 600 [28], 326 [16], 553 [32] 89 [33] |

The study performed by Fanton et al. [4] was a retrospective multicenter study that included 18,591 IVF cycles from three assisted reproductive technology centers in the USA. The large and diverse dataset ensured robust predictions of starting gonadotropin doses and enabled generalizability across clinical settings.

Hanassab et al. [15] performed a multicenter retrospective study involving 11 European IVF clinics and a total of 19,082 patients. The large-scale dataset allowed robust validation and generalization of the AI model across different patient populations and clinical protocols.

## 3.6 Use of Explainable AI (XAI)

Only three out of the reviewed papers explicitly addressed explainability and interpretability in their AI models, a critical aspect for ensuring trust and transparency in clinical applications. These studies leveraged tools like SHapley Additive exPlanations (SHAP) to elucidate decision-making processes, setting a precedent for integrating explainable AI (XAI) in reproductive medicine.

The study by Fanton et al. [4] prioritized interpretability in ML for optimizing the day of trigger during ovarian stimulation. The researchers employed a linear regression model to predict outcomes such as mature oocytes (MII), fertilized oocytes (2PNs), and usable blastocysts. This model allowed direct interpretability through standardized coefficients, providing clinicians insights into how specific follicle sizes contributed to recommendations. Recursive feature elimination ensured inclusion of the most relevant predictors, enhancing trust in the system's decisions. Unlike black-box models, this approach offered a transparent basis for its predictions, addressing potential clinician concerns and paving the way for broader adoption of AI in assisted reproduction.

The study of Hanassab et al. [15] utilized an ensemble-based explainable artificial intelligence (XAI) model to identify optimal follicle sizes on trigger day (TD) for maximizing mature oocytes, high-quality blastocysts, and live birth rates. Explainability was achieved through permutation importance and SHapley Additive exPlanations (SHAP) values, which identified follicles sized 13–18mm as most influential, with specific importance placed on 15–18mm. SHAP provided insights into how follicle sizes contributed to outcomes, while model validation via internal-external testing ensured robustness. The XAI approach supported a personalized, data-driven method for optimizing TD timing, moving beyond traditional reliance on lead follicle size, thereby enhancing clinician trust and potential patient outcomes.

The study by Fangfang et al. [14] employed SHapley Additive exPlanations (SHAP) to address explainability in their ML model for predicting reproductive outcomes after FET. SHAP provided both global interpretability, ranking feature importance such as the radiomics score, embryo grade, and endometrial thickness, and local interpretability, offering insights into individual predictions. These methods allowed clinicians to understand how specific features contributed to the model's decisions, bridging the gap between AI-driven insights and clinical trust. By making the model's decision-making process transparent, the study demonstrated the potential of explainable AI to enhance IVF treatment strategies effectively.

## 4 Discussion

This review highlights the transformative potential of AI in improving decision-making during ovarian stimulation for IVF. The analyzed studies demonstrate how AI can enhance critical processes, such as hormone dose adjustments, follicular monitoring, and optimal trigger timing. However, a key differentiation of this review with previously reported review papers is its explicit focus on applications where integration of medical imaging data with AI to support ovarian stimulation process assessment, setting it apart from previous reviews like Hariton et al. [8], which broadly examined AI applications in ovarian stimulation without a specific focus on imaging. Moreover, another differentiation of this review from previous ones is its focus on ovarian stimulation phase, which remains less explored in review studies compared to its widespread application in embryo assessment [34–36]. While AI combined with imaging is a well-established tool for embryo quality assessment, its use for analyzing ovarian or uterus imaging data before oocyte retrieval is significantly underrepresented, highlighting an important research gap identified by this review study.

Unlike Hariton et al., which primarily addressed clinical data and hormone optimization, this review underscores the limitations and opportunities of imaging data in IVF. Most of the studies analyzed in this review use imaging, primarily 2D ultrasound, for manual measurements such as follicle diameter and count from ovaries ultrasounds. In addition, a few studies used radiomics feature extraction from segmented endometrium using uterus imaging data. However, advanced imaging techniques and image analysis methods (e.g., 3D ultrasound or deep learning-based image analysis) are underrepresented. This gap represents a significant opportunity for innovation, where novel imaging and deep learning could unlock richer insights into follicular dynamics, endometrial receptivity, and their correlation with IVF outcomes particularly to support ovarian stimulation decision making process.

Additionally, imaging has often been relegated to supporting roles, focusing on basic parameters such as follicle size and count rather than leveraging the full potential of information which can be found in imaging data, such as follicle shape and patterns, texture, or volumetric changes.

This review also emphasizes the importance of explainable AI (XAI) in clinical settings, a topic that is insufficiently addressed in current literature but is pivotal for clinician trust and adoption. Among the studies analyzed, only three studies [14,15,20] explicitly incorporated XAI techniques, such as SHapley Additive exPlanations (SHAP), to provide transparency in decision-making. This enhances the review from Hariton et al., which does not address the explainability challenge in AI models. XAI methodologies are essential to bridge the gap between AI-driven insights and clinician confidence.

Finally, the predominance of single-center studies and small patient cohorts in current research limits the generalizability of AI applications. Multicenter studies, as seen in the works of Fanton et al. [4] and Hanassab et al. [15], demonstrate the advantages of larger, diverse datasets for building robust and generalizable AI models. This aspect aligns with the broader findings of Hariton et al. but is further contextualized here to emphasize the need for multicenter imaging data integration.

In conclusion, while this review reaffirms the transformative role of AI in ovarian stimulation, it uniquely advances the discussion by focusing on medical imaging integration and imaging data used, addressing its current underutilization, and advocating for future advancements in this area future research should explore underrepresented areas, such as applying advanced image analysis techniques to extract knowledge from follicle structure and uterus imaging to broaden the scope of AI applications and enhance outcomes in ovarian stimulation. In addition, explainable AI should be more utilized and diverse patient data from independent centers should be used to enhance explainability and generalizability of the developed AI methods. By addressing these gaps, AI can become a transformative tool, enabling more personalized and effective approaches to IVF treatments.